\documentclass[10pt,a4paper,twoside]{article}
\usepackage{epsfig}
\usepackage{baltlat6}
\usepackage{array}
\usepackage{here}
\begin{document}
\ \
\vspace{0.5mm}
\setcounter{page}{1}
\vspace{8mm}

\titlehead{Baltic Astronomy, vol.\, xx, yy 2011}

\titleb{The long-term spectroscopic misadventures of AG Dra with a nod toward V407 Cyg: Degenerates behaving badly} 

\begin{authorl}

\authorb{S. N. Shore,}{1} 
\authorb{K. Genovali,}{1,2}
\authorb{G. M. Wahlgren}{3,4}

\end{authorl}

\begin{addressl}
\addressb{1}{Dipartimento di Fisica "Enrico Fermi", Universit\`a di Pisa\\  largo B. Pontecorvo 3, Pisa 56127, Italia; shore@df.unipi.it}
\addressb{2}{Dipartimento di Fisica, Universit\`a di Roma Tor Vergata\\ Via della Ricerca Scientifica, 1, I-00133 
Roma, Italia; katia.genovali@roma2.infn.it}
\addressb{3}{Department of Physics, Catholic University of America\\
 620 Michigan Ave., N.E.,  Washington, DC 20064  USA; glenn.m.wahlgren@nasa.gov}
\addressb{4}{NASA Goddard Space Flight Center, Code 667,  Greenbelt, MD 20771 USA}
\end{addressl}

\submitb{Received: }

\begin{summary} 
We present some results of an ongoing study of the long-term spectroscopic variations of AG Dra, a prototypical eruptive symbiotic system.  We discuss the effects of the environment and orbital modulation in this system and some of the physical processes revealed by a comparison with the nova outburst of the symbiotic-like recurrent nova V407 Cyg 2010.
\end{summary}


\resthead{AG Dra spectroscopic history}
{S. N. Shore et al.}

\sectionb{1}{INTRODUCTION}

The symbiotic binary AG Dra certainly needs no introduction to readers of these proceedings and we refer to Leedj\"arv's report in these proceedings for the relevant background and the excellent introductions in Leedj\"arv et al. (2004), Viotti et al. (2007), G\'alis et al. (2007), and Gonz\'alez-Riestra et al. (2008).  Photometric monitoring of this system extends now over nearly a century with the highest precision spanning more than 40 years from both amateurs and professional astronomers (see Munari, these proceedings, and Skopal et al. (2009)).   We concentrate, instead, on the line profiles and flux measurements.   We have used new and archival optical grating, grism, and echelle spectra from Asiago (Pennar and Ekar), Loiano, the Telescopio Nazionale Galileo (TNG), Ond\v{r}ejov, and the Nordic Optical Telescope (NOT).   Part of these data have already been published (Shore et al. 2010) and we refer the reader to that paper for details covering the 2006-2009 part of the observations.  The full details of the data set are available in Genovali (2010).  Here we outline some of the results from our study of AG Dra that began with the major outburst of 2006-2008 and has since expanded to cover the last 30 years  (Genovali et al. in preparation).   

\sectionb{2}{EMISSION LINE FLUX VARIATIONS}

We begin with two summary plots.  Figure 1 is the variation of the H$\alpha$ through H$\delta$ line equivalent widths, corrected for the continuum variations, from the major outburst of 2006 of AG Dra.    Figure 2 shows the relative He I, He II 4686\AA, and O VI 6825\AA\  variability during the 2006 outburst.  We draw the reader's attention to the bifurcation in the last panel that may distinguish the hot and cool outbursts; those from the start of the 2006 eruption are the upper branch for He I.  As noted by Gonz\'alez-Riestra et al. (2008) and Skopal et al. (2009), there is a saturation in the He II emission level during the major eruptions.  This could be due to the increased optical depth of the wind of the hot component.  Viotti et al. (2007) have already conjectured that a burst may have two stages depending on its intensity:  a hot event may, if it continues long enough, become a cool outburst.

\begin{figure}[!tH]
\vbox{
\centerline{\psfig{figure=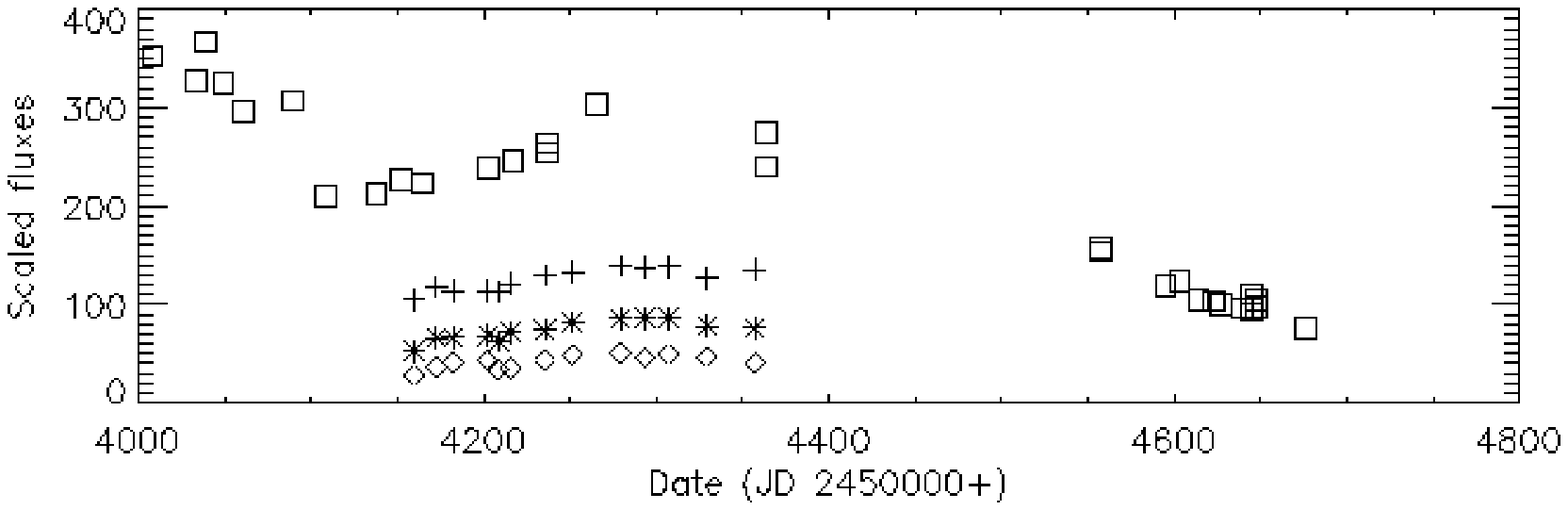,width=110mm,angle=0,clip=}}
\vspace{1mm}
\captionb{1}
{The Balmer line  variations during the 2006-2008 outburst of AG Dra, corrected for variations in the continuum flux as described in Shore et al. (2010).   Square: 
H$\alpha$; cross, H$\beta$; asterisk: H$\gamma$; diamond, H$\delta$.  The outburst covered the orbital interval 5.$^p$88-7.$^p$15 with the peak fortuitously occurring near superior conjunction of the giant.}
}
\end{figure}

\begin{figure}[!tH]
\vbox{
\centerline{\psfig{figure=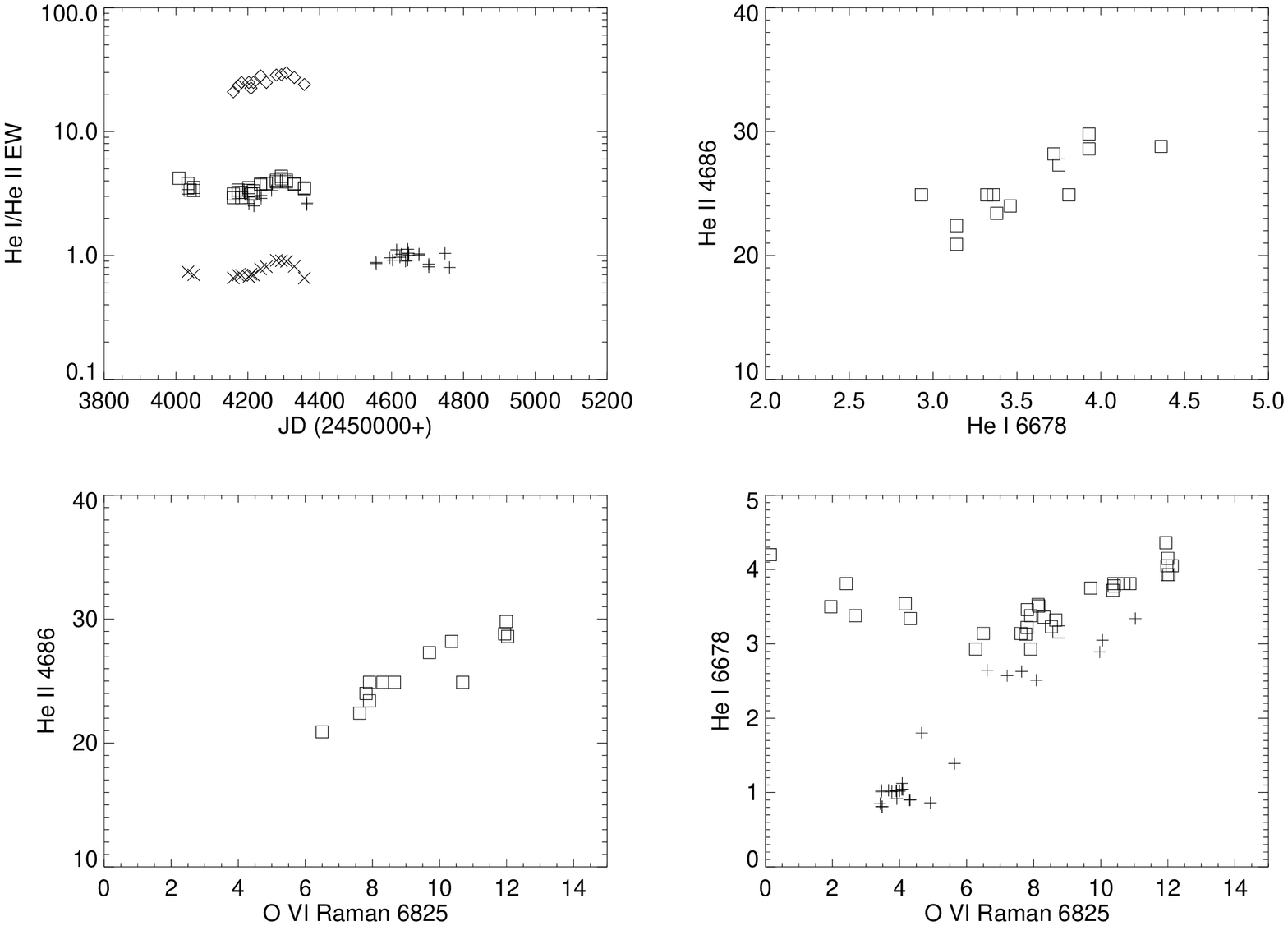,width=90mm,angle=90,clip=}}
\vspace{1mm}
\captionb{2}
{Variations of the He I, He II, and O VI Raman 6825\AA\ line during the 2006-2008 eruption from the Loiano and Ond\v{r}ejov spectra.  Top left: He I 6678\AA\  (square). He I 7065\AA\ (cross), and He II 4686\AA\ (diamond) variations in time.  Top right, He I 6678\AA\ vs.He II 4686\AA.  Bottom left,  He II 4686\AA\ vs. O VI Raman 6825\AA; bottom right: He I 6678\AA\ vs. O VI Raman 6825\AA.  Note, in the last plot, the bifurcation.}
}
\end{figure}

Passing to the long haul, we show in Fig. 3 the variations in the equivalent widths of the main emission features and the UBV photometry for about two decades.\footnote{ The displayed data are from Leedj\"arv et al. (2004) (star), Tomova \& Tomov (1999) (pentagon)  and spectra from Loiano (plus), Asiago/Ekar (asterisk), Catania (diamond), TNG (cross), Asiago/Pennar (filled circle).  In the lower panel: U (filled circle), B (plus), and V (circle) from Skopal and collaborators and the AAVSO.}   It is an impressively rich phenomenology (see also G\'alis et al. 2007).  Informed by the dense coverage of the 2006 outburst, going backwards in time to the major outburst of 1995, it appears that the pattern of Raman 6825\AA\ feature variation was the same.  A steep rise from what was likely the disappearance of the line.  A relatively minor outburst, about three years later, shows the same pattern.   Leedj\"arv reported at the meeting that the same thing appears to be happening again with the most recent (2010) minor outburst.  Since the variations of this line are directly linked to the down-conversion of O VI 1032\AA\ photons, there are two possible, likely linked inferences: the column density in the wind around the white dwarf (WD) could have changed and/or the intensity of the resonance lines could have changed.  The first implies a variability of the mass loss rate from the red giant (RG) or a change in the ambient density around the WD.  The second, instead, is something intrinsic to the WD mass loss, a change in the optical depth of the O VI lines implying a variation in the wind from the degenerate component.   Two close peaks around day 1000 (Fig. 3) show the same Raman line variation eventually reaching the same maximum.  If these were so-called {\it hot} outbursts, the O VI line would have turned less optically thick than in the {\it cool} outbursts, and the feature would not have disappeared.  If, instead, the WD mass loss was  higher, and the covering factor complete, then the increase in the WD wind could have been so large that the O VI 1032, 1036\AA\ was smuthered with a consequent vanishing of the optical features (cf. Schmid (1995) and Harries \& Howarth (1997)  for a discussion).

\begin{figure}[!tH]
\vbox{
\centerline{\psfig{figure=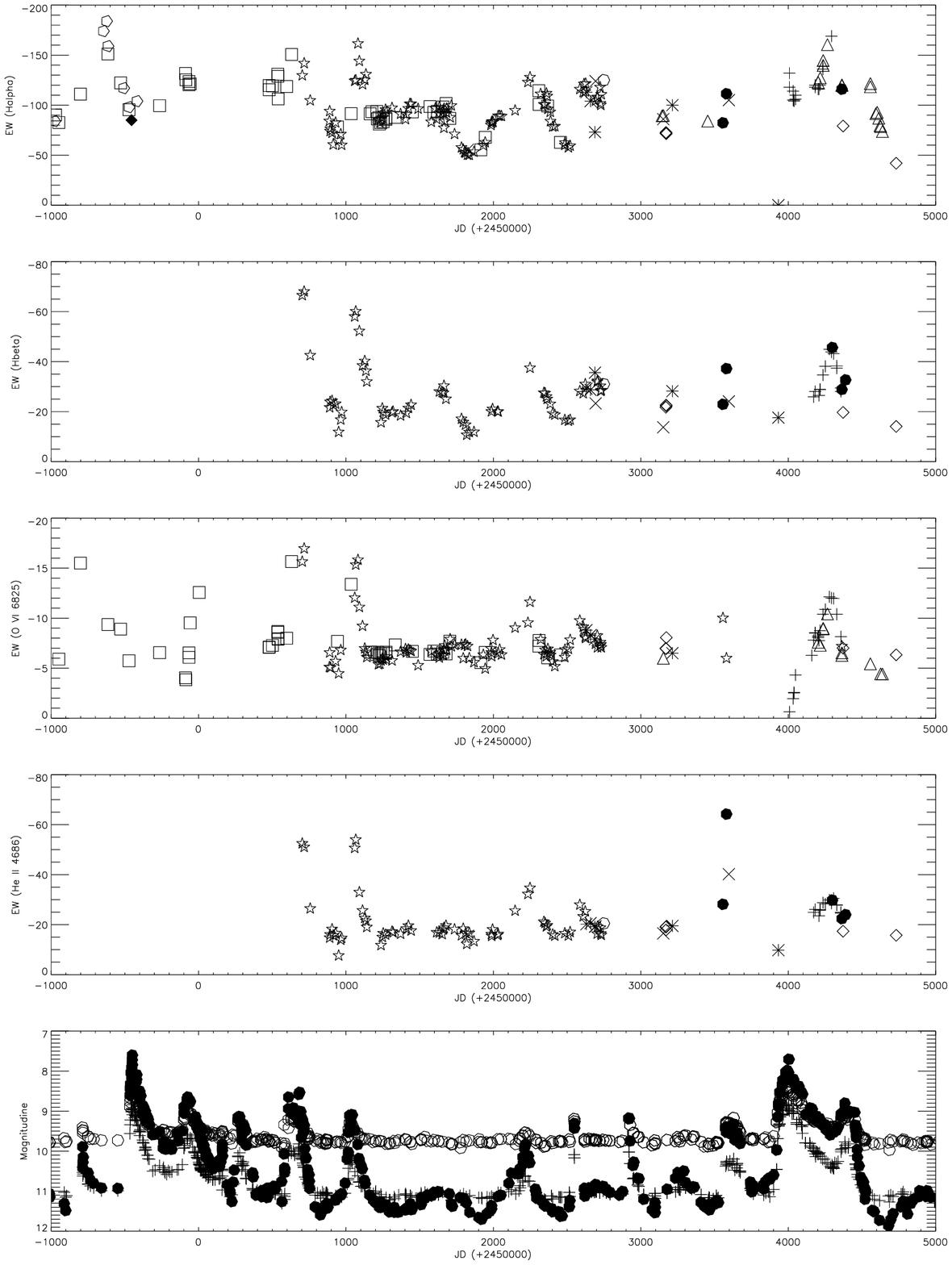,width=130mm,angle=0,clip=}}
\vspace{1mm}
\captionb{3}
{The long term spectroscopic history of AG Dra.  Equivalent width variations of  (top to bottom) H$\alpha$, H$\beta$, O VI Raman 6825\AA, and He II 4686\AA\ emission lines, corrected for continuum variations, together with the UBV photometry. }
}
\end{figure}

The 2006 outburst confirmed what had, implicitly, already been reported in the literature, that the Raman lines can disappear (Munari et al. 2009).  Having been seen on only one occasion, however, and lacking dense coverage it was not obvious that this is a generic feature of a certain magnitude of event.  Instead, with 20-20 hindsight, it seems that the 6825\AA\ feature flux curve observed during the last major outburst is actually typical, the slow recovery of the line relative to its sharp initial drop and its subsequent return to the quiescent state being recognizable in earlier events.  The interval of disappearance was brief, about a week, and due to weather, season, and  scheduling,  other similar events might have been (barely) missed.  The rise from minimum line strength is similar for all such large U outbursts.  An important feature of the long term variations is the appearance of a sort of saturation level for the emission line strengths, in particular He II 4686\AA\ as a function of the U magnitude.   From Leedj\"arev et al. (2004), it was already known that the emission varies in phase with the near UV but the relation was a simple power law up to U$\approx 9^m$.  At brighter levels, the strict increase breaks down with a large dispersion, greater for He I 6678\AA\ than for He II 4686\AA.  The O VI Raman 6825\AA\ feature actually peaks at U $\approx$10 and then declines by about a factor of two for increasing brightness.  The maximum equivalent width reached by the Raman 6825\AA\ line is -15\AA\ that is, however, rarely reached in the historical data.

The optical He II lines, formed by recombination in the circum-WD region, are useful, but not unique, proxy measures of the state of the hot source.  The recombination time is density dependent and can be of the same order as the interval between activity phases.  It should be recalled that the environment is a mess given the many dynamical and radiative processes acting simultaneously.  These lines convolve the ionization source variations with the response function of the environment, an unknown given the uncertainties regarding the mechanism producing the ultraviolet and X-ray variations.  The Raman features are, instead, instantaneous measures of the FUV since they are optically thin and produced {\it only} by scattering.  Using this, it is interesting that one recovers the same classification of {\it hot} and {\it cool} outbursts obtained from the direct measurement of the X-rays with ROSAT and XMM analyzed by Gonz\'alez-Riestra et al. (2008) and Skopal et al. (2009).  Those events during which the Raman line fell below the quiescent floor and the U magnitude increased by the largest amount were judged cool, while the the two events that saw the line increase from the floor level and return to it after a short time  were hot.  The other intriguing feature is that during the hot outbursts the Raman line reaches a systematically greater flux.  

The H$^+$ region of the WD is always ionization bounded, we are not dealing here with an interstellar cloud, but it is dense and not without intrinsic line opacity that can affect the spectrum formation at other wavelengths.  For example, changes in the Lyman series optical depths, that can be caused by variations in the FUV luminosity of the hot source or a change in the absorbing column density, can only be extracted by detailed modeling of the line formation.  Decreasing the density increases the ionized volume and decreases the Lyman series optical depth, changing the optical Balmer line intensities and profiles.  Changing the mass loss rate from the hot star displaces the location of the stagnation point of the wind-wind shock and also may change the spectrum of hard photons available to further ionize the environment.   This requires self-consistent, time dependent photoionization and hydrodynamic models of the sort described by Folini et al. (1998, 2003).   

Like the Raman events, there seems to be a characteristic  time development of the U outbursts but these also indicate different natures.   A change in the covering factor and/or the column density without an accompanying change in the luminosity of the hot source will produce only flux redistribution and the bolometric luminosity should remain constant.  This is well known from the post-explosion stage of classical novae and is not far from correct for LBVs.  In such cases, an increase in the mass loss rate from the hot star could have the same effect.  Instead, as suggested by Sokoloski et al. (2006), if the outbursts are actually mini-TNR events (see Starrfield, these proceedings), then the bolometric luminosity should vary.  At least the major event of 2006 appears to be the latter case.  This is a further way to separate the sources of the variations. 

\sectionb{3}{ORBITAL MODULATION OF STRONG EMISSION LINES}

Line profile variations are direct probes of the environmental effects along the line of sight.  To illustrate the complexity of the AG Dra phenomenology, we show some sample phased variations of the H$\alpha$ and helium lines.  Figure 4 shows the He I 6678\AA\ and O VI 6825\AA\ scaled flux variations plotted on the Fekel et al. orbital solution.  No distinction is made in this plot between phases of activity from JD 49030 to JD 54639 (Ond\v{r}ejov and Loiano spectra only).  For the Raman feature, the minimum line strength ($<0.3$\AA) from the 2006 outburst happens to have occurred at orbital phase zero.  This plot also suggests the presence of multiple periodicities in the line variations.  While the Raman line shows little orbital modulation, there is a clear indication of such in the He I variations.  Further data cuts, based on the activity state, show that the line fluxes from pre-selected quiescent stages are those with the strongest orbital modulation, as can be gleaned from Fig. 3. 

\begin{figure}[!tH]
\vbox{
\centerline{\psfig{figure=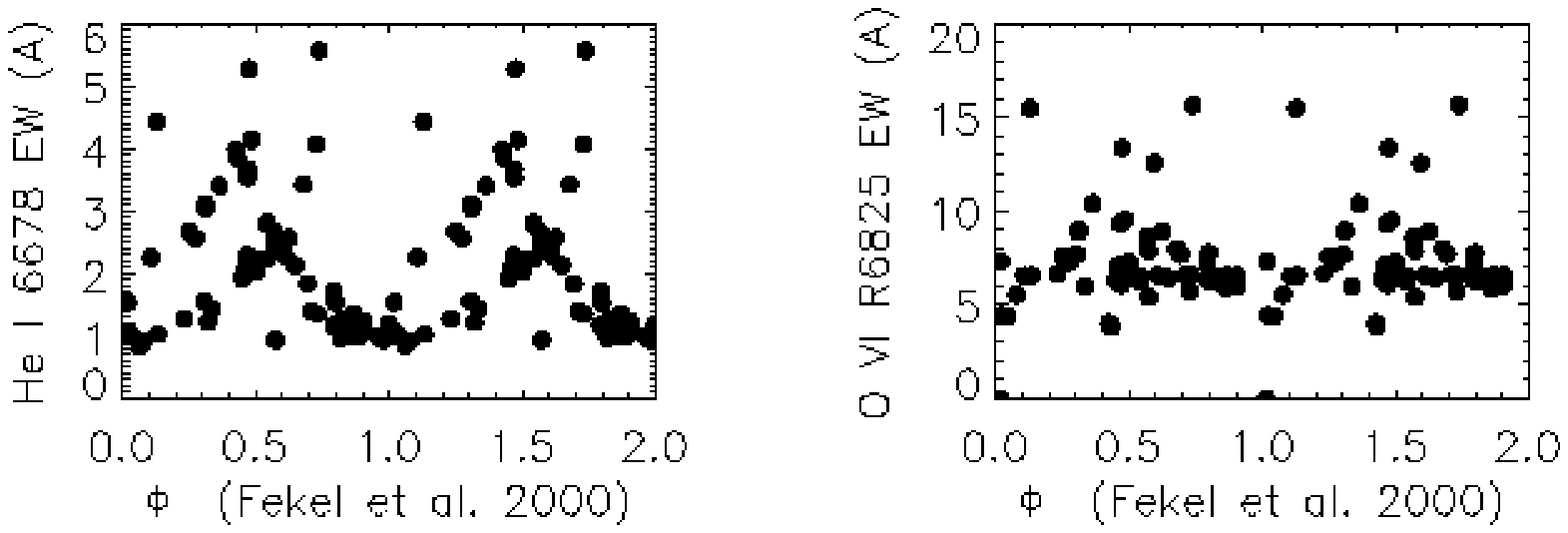,width=120mm,angle=0,clip=}}
\vspace{1mm}
\captionb{4}
{The equivalent width variations of the  O VI Raman 6825\AA, and He I 6678\AA\ emission lines in AG Dra, corrected for continuum variations, phased on the orbital solution by Fekel et al. (2000).  See text for details.}
}
\end{figure}

The Raman features are immune to self-absorption, that alters the permitted Balmer and helium profiles, and Rayleigh scattering that affects the UV lines. Electron scattering should not be important since it is local to the source, not to the conversion region.  Instead, they should be sensitive to changes in the line of sight column densities of neutral hydrogen.\footnote{ A similar effect was discussed for the O I] 1641\AA\ line in symbiotics by Shore \& Wahlgren (2010) as a probe of the neutral wind of the RG.  In the absence of available UV measurements, the [O I] 6300\AA\ line, formed from the same transition cascade, may be an appropriate substitute.}  The modulation of the ratio of the two optical O VI Raman features may, consequently, be indicative of structural changes in the wind resulting from the hydrodynamic interaction of the winds from the two components (see Fig. 4 in Shore et al. 2010).   The minimum in the line ratio happened in the interval between the two U maxima of the outburst and is, perhaps coincidentally, approximately at inferior conjunction of the WD.  We urge observers to include, whenever possible, both features in future spectroscopic studies.  In Figs. 5a,b and 6a,b we show sample high resolution line profiles at a few phases to illustrate the environmental effects.

\begin{figure}[!tH]
\vbox{
\centerline{\psfig{figure=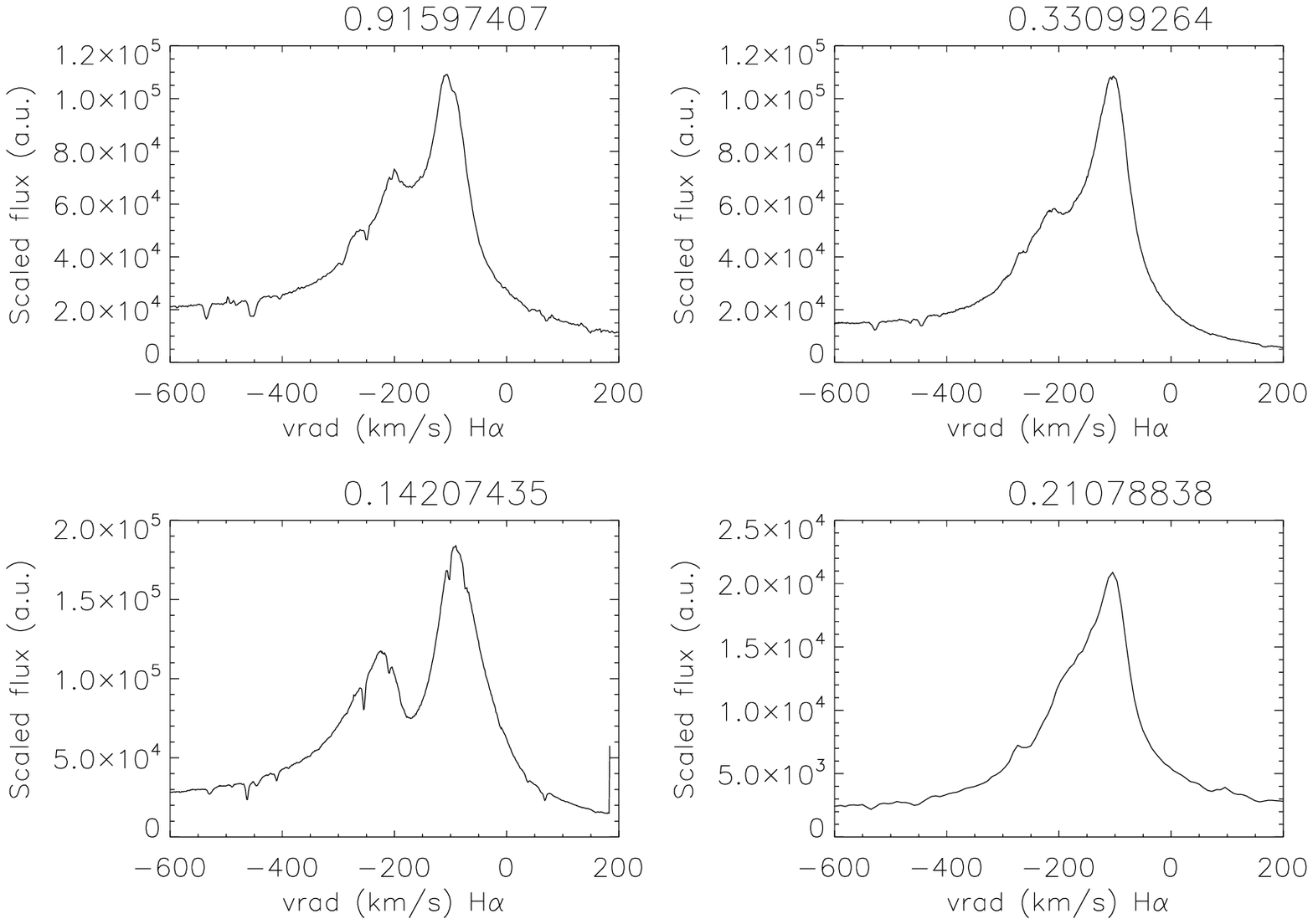,width=110mm,angle=0,clip=}}
\vspace{1mm}
\captionb{5a}
{Sample profiles for H$\alpha$ in AG Dra from the TNG spectra at several orbital phases.  The dates are JD 51826, 53151, 53589, and 54731. The phases are according to  the ephemeris of Fekel et al., (2000) with $\phi = 0^p$ is RG inferior conjunction. }
}
\end{figure}

\begin{figure}[!tH]
\vbox{
\centerline{\psfig{figure=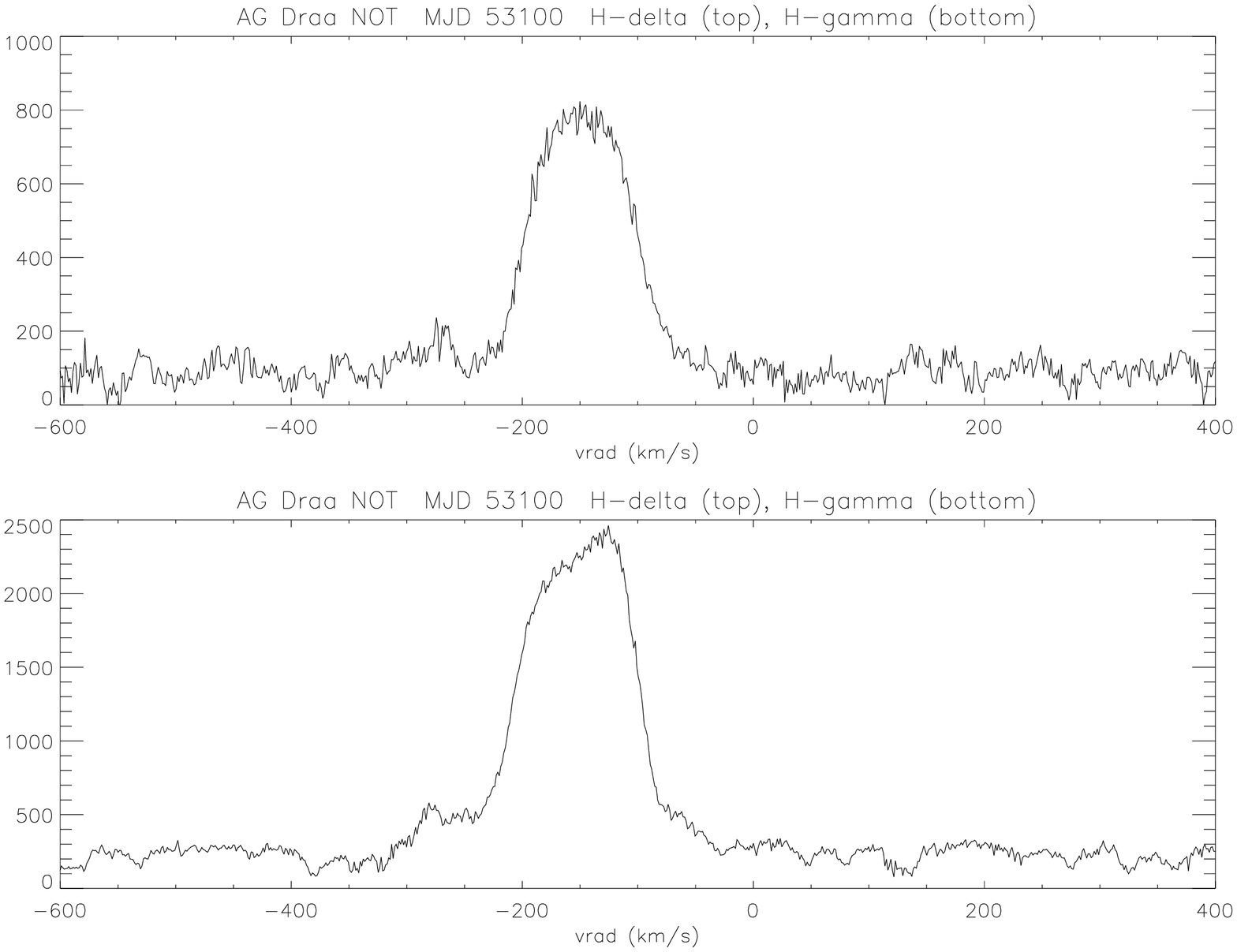,width=100mm,angle=0,clip=}}
\vspace{1mm}
\captionb{5b}
{The Balmer H$\gamma$ and H$\delta$ line profiles from the NOT spectrum on JD 53100, at orbital phase 0.$^p$23.}
}
\end{figure}

\begin{figure}[!tH]
\vbox{
\centerline{\psfig{figure=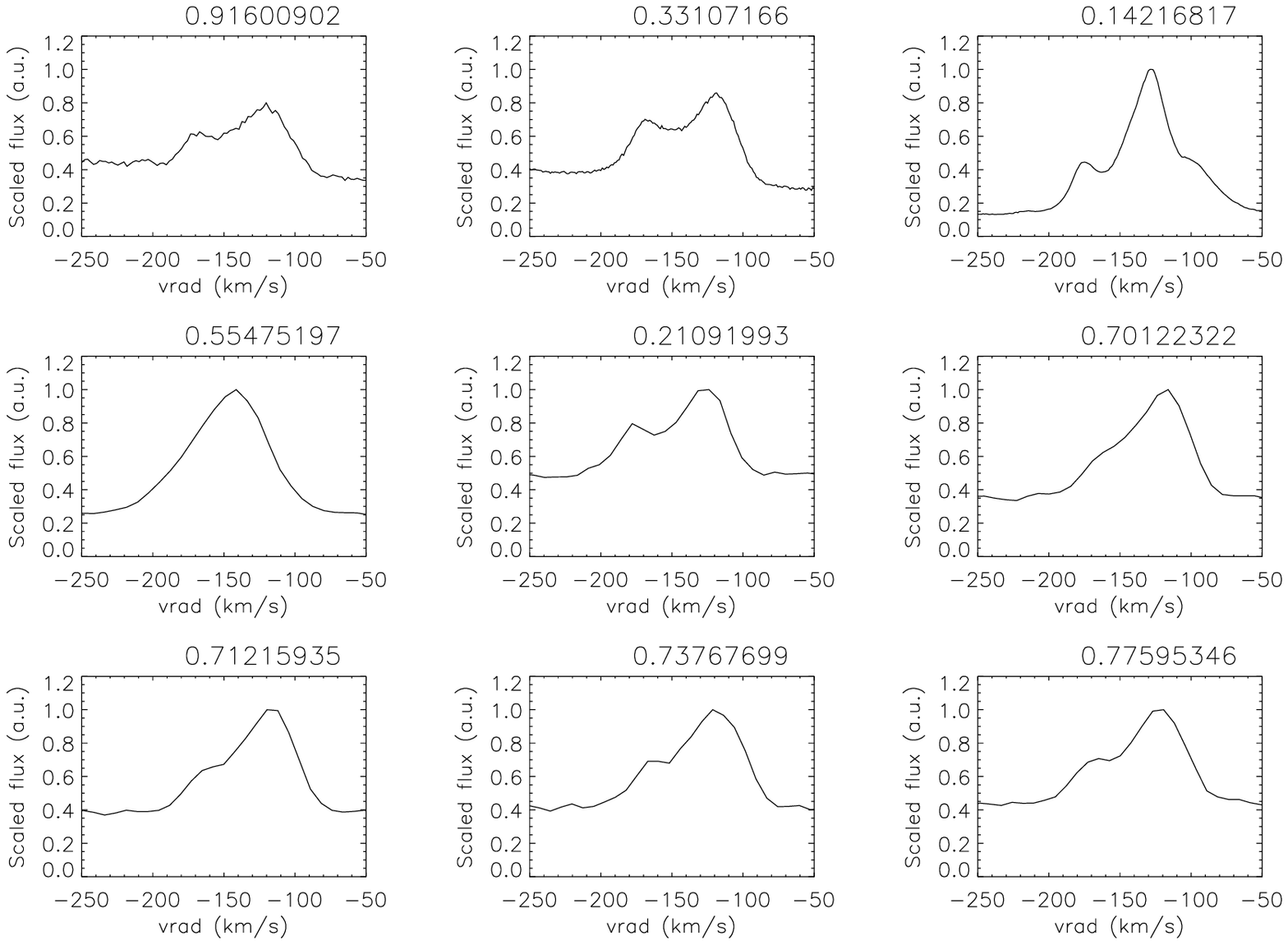,width=130mm,angle=0,clip=}}
\vspace{1mm}
\captionb{6a}
{Examples of the He I 5875\AA\ line profiles in AG Dra as a function of orbital phase from Catania, TNG, and Asiago spectra.  The dispersions were 0.02\AA\ px$^{-1}$ and 0.15\AA\ px$^{-1}$.  The time period is from JD 51826 to JD 55076 covering eight orbits. }
}
\end{figure}

\begin{figure}[!tH]
\vbox{
\centerline{\psfig{figure=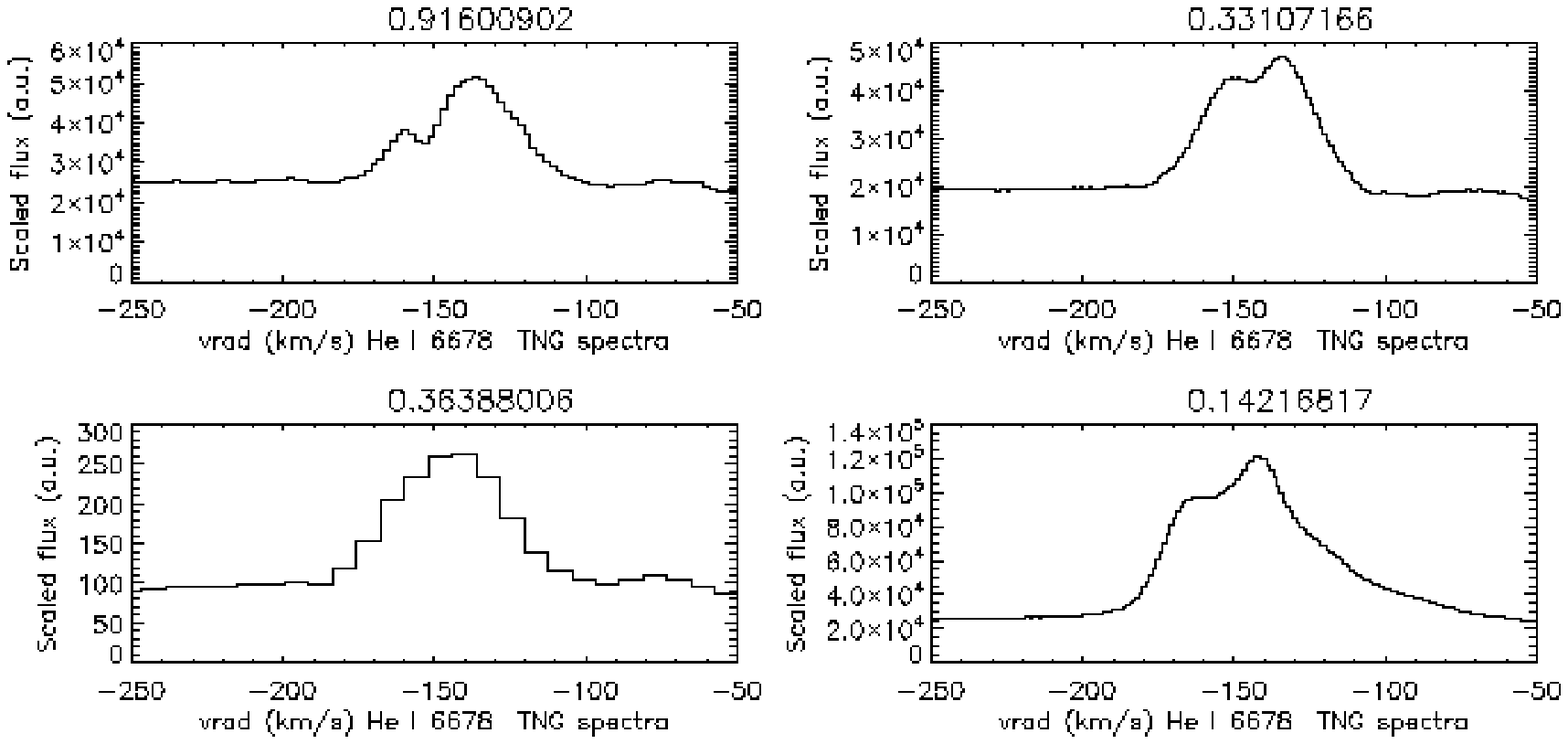,width=130mm,angle=0,clip=}}
\vspace{1mm}
\captionb{6b}
{Examples of the structure on the He I 6678\AA\ singlet line profiles; as in Fig. 6a, from TNG and Asiago. }
}
\end{figure}

\begin{figure}[!tH]
\vbox{
\centerline{\psfig{figure=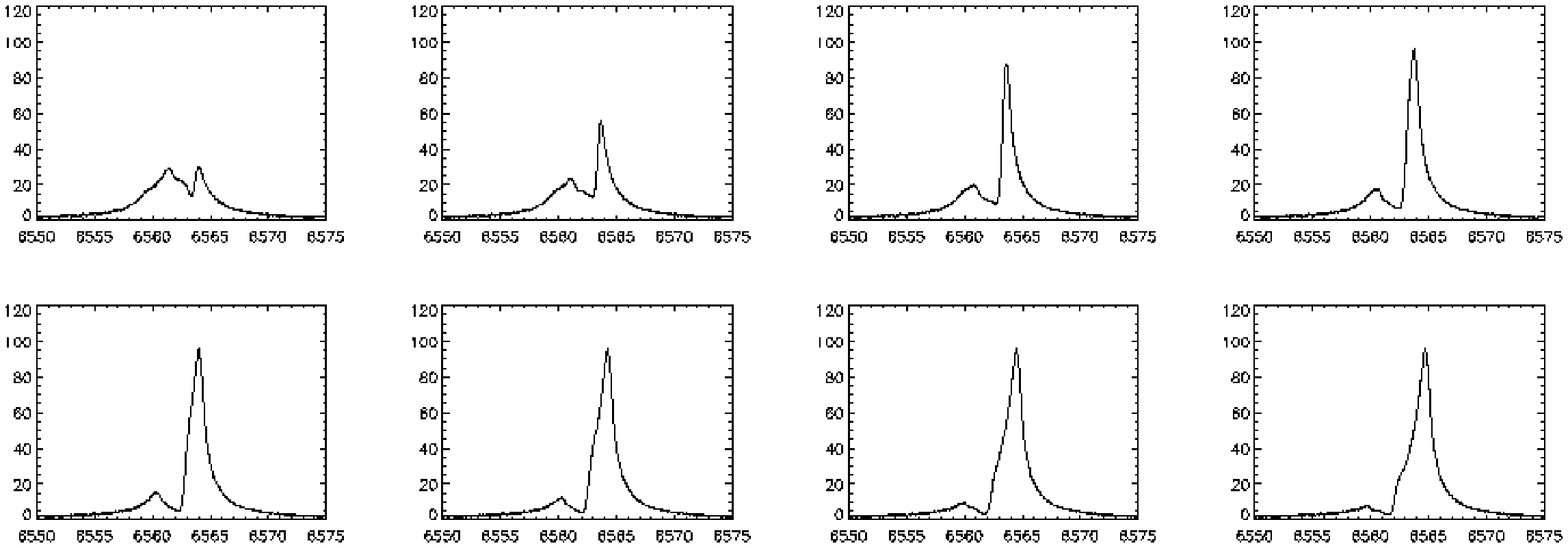,width=50mm,angle=90,clip=}}
\vspace{1mm}
\captionb{7}
{Model line profile variations with orbital phase.  A synthetic H$\alpha$ line, assumed to be a gaussian, was absorbed in the wind of the RG assuming a terminal velocity of 30 km s$^{-1}$ and a simple $\beta$-law (see text).  The mass ratio and orbital period approximately match BX Mon.  Differential displacement (center-of-mass frame) produces the variability of the absorption feature.  The emission line intensity was kept constant.}
}
\end{figure}

The He I line profiles show a nebular structure, the He II 4686, 5412\AA\ lines show, instead, a comparatively single peaked profile  that remains at almost constant radial velocity, about -200 km s$^{-1}$, but whose symmetry changes.   The variations of the profile of the 1640\AA\ line, studied by Gonz\'alez-Riestra et al. (2008), are more extreme in displaying extended wings with maximum velocities of 2000 km s$^{-1}$ that do not appear to show orbital modulation.

Figure 7 shows a simple model of the line profile variations for H$\alpha$ as a function of orbital phase (see Dumm et al. (1998) for an application to  BX Mon).  The emission line from the ionized region around the hot star, assumed to be a gaussian for simplicity, is observed with orbital radial velocity shifts through the extended neutral wind of the RG that has a $\beta$-type velocity law.  In this case, a velocity amplitude of order 10 km s$^{-1}$ for the RG and a mass ratio of 3:1 (RG:WD) were assumed to simulate BX Mon (see Dumm et al. (1998)).  No quantitative information is intended here, just a warning  that even in the visual part of the spectrum, environmental line absorption effects that are well known to affect ultraviolet line profile studies can produce significant changes in the profiles.   It then suffices to recall that in the real systems, the interactions of the winds of the two components, complex structures within the orbital plane, non-spherical outflows, and time dependent variations, combine to complicate the quantitative analyses of such variations.  In AG Dra, for instance, the He II 1640\AA\ emission line intensity variations and profiles are clearly affected by both intrinsic changes {\it and} orbital modulation by the UV Fe-curtain and detailed models are needed to separate the various contributions.

\sectionb{4}{LESSONS LEARNED FROM THE 2010 NOVA ERUPTION OF V407 CYG}

One of the many things learned from the 2010 eruption of V407 Cyg, a symbiotic that showed all the symptoms in this event of a symbiotic-like recurrent of the RS Oph variety (Munari et al. 2011, Shore et al. 2011a,b), is that the presence of broad wings on the emission lines, especially intrinsically asymmetric profiles, can also have hydrodynamic origins aside from steady outflows.  The interactions in the environments of these systems are complex and very time dependent, especially in the lower density parts of the circumstellar medium, as shown by two and three dimensional simulations.  We show some examples of the late-time Balmer line profiles that allow a determination of the individual regions and their reaction to the expanding shock and photoionization changes caused by the post-eruption emission from the WD.  The lines are asymmetric, a characteristic feature of an off-center shock propagating in the density gradient of the RG wind.  Such profiles have, intriguingly, been observed on He II 1640\AA\ in AG Dra and other symbiotics in outburst.  The subsequent passage of the shock and the onset of wind recombination renders the profiles more symmetric with time, much line the NOT observation of H$\gamma$ and H$\delta$ in AG Dra (Fig. 5b).

\begin{figure}[!tH]
\vbox{
\centerline{\psfig{figure=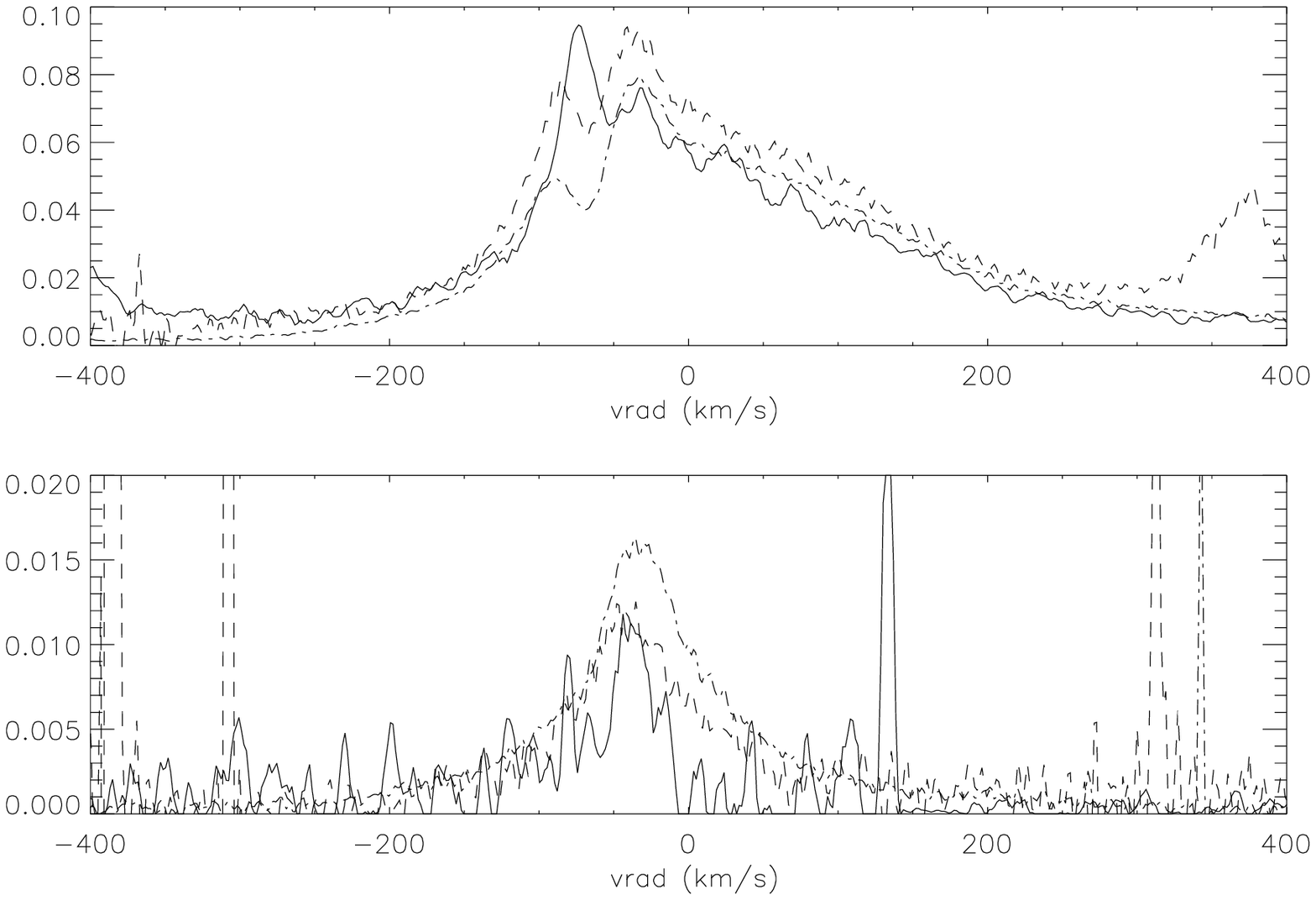,width=90mm,angle=0,clip=}}
\vspace{1mm}
\captionb{8}
{The Balmer line profile in the symbiotic-like recurrent nova V407 Cyg on 2010 Jul. 15 (top) and 2011 Aug. 21 (bottom) (date of outburst 2010 Mar. 10) from the NOT (0.02\AA\ px$^{-1}$).  The profiles are H$\beta$ (dot-dash), H$\gamma$ (dash), and H$\delta$ (solid).  The maximum shock velocity was $>$3000 km s$^{-1}$.  The stellar velocity is -54 km s$^{-1}$, note the wind absorption at an expansion velocity of about 10 km s$^{-1}$  that becomes optically thin as the outburst progresses.  The broad wings are due to a shock from the ejecta and subsequent outflow in the Mira (see discussion and Shore et al. 2011a,b).  }
}
\end{figure}

\thanks{We thank T. Iijima, C. Rossi, and R. Viotti  for providing some of the spectra used in this study.  SNS thanks the organizers for their kind invitation and patience, and D. Folini, J. Jos\'e, P. Koubsky,  L. Leedj\"arv, J. Mikolajewska, K. Mukai, C. Rossi, A. Siviero, A. Skopal, J. Sokoloski, S. Starrfield, R. Viotti, and R. Walder for stimulating  discussions.  Extensive use has been made during this project of the Simbad, ADS, and MAST databases.  We also thank the AAVSO for providing archival photometry.  GMW acknowledges support from NASA grant NNG06GJ29G.}

\References

\refb Dumm, T., Muerset, U., Nussbaumer, H.  et al. 1998, A\&A, 336, 637

\refb Fekel, F. C., Hinkle, K. H., Joyce, R. R., \& Skrutskie, M. F. 2000, AJ, 120, 3255

\refb Folini, D. \& Walder, R. 1998, AGAb, 14, 108

\refb Folini, D., Walder, R., Psarros, M. \& Desboeufs, A. 2003, ASPC, 288, 433

\refb G\'alis, R., Hric, L., Leedj\"arv, L., \& \v{S}uhada, R 2007,  IAUS, 240, 121                  

\refb Genovali, K. 2010, MSc Thesis - Physics, Univ. of Pisa\\ (URL:  http://etd.adm.unipi.it/theses/available/etd-05052010-002805/)

\refb Gonz‡lez-Riestra, R., Viotti, R., Iijima, T. \& Greiner, J. 1999, A\&A, 347, 478

\refb Gonz\'alez-Riestra, R.; Viotti, R. F.; Iijima, T. . et al. 2008, A\&A, 481, 725

\refb Harries, T. J. \& Howarth, I. D. 1997, A\&AS, 121, 14

\refb Leedj\"arv, L., Burmeister, M., Mikolajewski, M. et al. 2004, A\&A, 415, 273

\refb Munari, U., Joshi, V. H., Ashok, N. M.  et al. 2011, MNRAS, 410, L52
\refb Munari, U., Siviero, A., Ochner, P. et al. 2009, PASP, 121, 1070

\refb Schmid H.M., 1995, MNRAS 275, 227

\refb Shore, S. N. \& Wahlgren, G. M  2010, A\&A, 515, A108

\refb Shore, S. N., Wahlgren, G. M., Genovali, K.  et al 2010, A\&A, 510, A70

\refb Shore, S. N., Wahlgren, G. M., Augusteijn, T. et al. 2011a, A\&A, 527, A98

\refb Shore, S. N., Wahlgren, G. M., Augusteijn, T.  et al. 2011b, submitted to A\&A

\refb Skopal, A., Seker\'‡\v{s}, M., Gonz\'alez-Riestra, R., \& Viotti, R. F. 2009, A\&A, 507, 1531

\refb Skopal, A., Va\v{n}ko, M., Pribulla, T.,   et al. 2007, AN, 328, 909

\refb Sokoloski, J. L., Kenyon, S. J., Espey, B. R.  et al. 2006, ApJ, 636, 1002

\refb Tomova, M. T., \& Tomov, N. A. 1999, A\&A, 347, 151

\refb Viotti, R. F., Friedjung, M., Gonz\,alez-Riestra, R., et al. 2007, BaltA, 16, 20

\end{document}